\documentclass[%
 reprint,
 amsmath,amssymb,
 aps,
]{revtex4-1}

\usepackage{graphicx}
\usepackage{dcolumn}
\usepackage{bm}

\begin{document}

\preprint{APS/123-QED}

\title{Tachyon Mimetic Inflation as an Instabilities-Free Model}
\author{Narges Rashidi}\homepage{n.rashidi@umz.ac.ir}

\author{Kourosh Nozari} \homepage{knozari@umz.ac.ir}

\affiliation{Department of Theoretical Physics, Faculty of Basic Sciences,\\
University of Mazandaran,\\
P. O. Box 47416-95447, Babolsar, IRAN}

\begin{abstract}
We consider the mimetic tachyon model in the lagrange multiplier
approach. We study both the linear and non-linear perturbations and
find the perturbation and non-gaussianity parameters in this setup.
By adopting two types of the scale factor as the power-law
($a=a_{0}\,t^{n}$) and intermediate ($a=a_{0}\exp(bt^{\beta})$)
scale factors, we perform a numerical analysis on the model which is
based on Planck2018 TT, TE, EE+lowE+lensing +BAO +BK14 and
Planck2018 TTT, EEE, TTE and EET data sets. We show that the mimetic
tachyon model with both the power-law and intermediate scale
factors, in some ranges of its parameter space is instabilities-free
and observationally viable. The power-law mimetic tachyon model with
$26.3<n<33.0$ and the intermediate mimetic tachyon model with
$0.116<\beta<0.130$ are consistent with observational data and free
of the ghost and gradient
instabilities.
\end{abstract}

\maketitle

\section{Introduction}

Although a canonical scalar field, slowly rolling its flat
potential, gives a simple model of inflation which solves some
problems of the standard model of cosmology, it predicts the scale
invariant, adiabatic and almost gaussian dominant modes of
the primordial perturbations  (precisely, this model predicts small,
still non-vanishing primordial non-Gaussianity)
~\cite{Gut81,Lin82,Alb82,Lin90,Lid00a,Lid97,Rio02,Lyt09,Mal03}.
However, some extended models of inflation, predicting the
non-Gaussian distributed perturbations, have attracted a lot of
attentions~\cite{Mal03,pl19,Bar04,Che10,Noj11,Fel11a,Fel11b,Noz15,Noz15b,Noz16,Noj17,Noz17,Noz18b,Ras20}.
One way to get a model with non-Gaussian distributed perturbations,
is to consider the non-canonical scalar field such as DBI or tachyon
fields~\cite{Sil04,Che07,Ali04,Che05,Sen99,Sen02a,Sen02b,Noz13a,Noz13b,Ras18,Noz19,Ras19b}.
In this paper, we focus on the tachyon scalar field. It is possible
to consider this scalar field, which is associated with D-branes in
string theory~\cite{Sen99,Sen02a,Sen02b}, as responsible for both
the early time inflation~\cite{Sam02,Noz13a} and late time
acceleration~\cite{Cop05,Pad02,Noz13a} .

On the other hand, in 2003 Chamseddine and Mukhanov have proposed a
new approach to the General Relativity in which a non-dynamical
scalar field ($\phi$) relates the physical metric ($g_{\mu\nu}$) to
an auxiliary metric ($\tilde{g}_{\mu\nu}$) as follows~\cite{Cham13}
\begin{eqnarray}\label{eq1}
g_{\mu\nu}= -\tilde{g}^{\alpha\beta}\,\phi_{,\alpha}\,\phi_{,\beta}
\,\tilde{g}_{\mu\nu}\,.
\end{eqnarray}
This proposal, called mimetic gravity, has this important property
that respects the conformal symmetry as an internal degree of
freedom~\cite{Cham13}. Also, the definition (\ref{eq1}) ensures that
by performing a weyl transformation on the auxiliary metric, the
physical metric remains invariant. In the mimetic gravity, there is
the following constraint
\begin{equation}\label{eq2}
g^{\mu\nu}\phi_{,\mu}\phi_{,\nu}=-1\,,
\end{equation}
obtained from equation (\ref{eq1}). The contribution of the matter
fields coupled to $g_{\mu\nu}$ in the action of the mimetic gravity,
leads to an extra term in the Einstein's field equations
corresponding to $a^{-3}$. This extra term in the field equations,
which mimics the matter component, is considered as a source of the
dark matter~\cite{Cham13}. The mimetic gravity scenario can be
studied in another approach by considering the lagrange multipliers
in the action of the theory, as proposed in
Refs.~\cite{Arr15,Gol13,Ham13}. In Ref.~\cite{Bar13}, the authors
have studied some models of the mimetic gravity which are
ghost-free. Also, considering a potential term for the mimetic field
leads to some interesting cosmological results. This case has been
discussed in ~\cite{Cham14}. Several extension of the mimetic
gravity have been studied by authors, such as the braneworld mimetic
scenario~\cite{Sad17}, non-minimal coupling in the mimetic
model~\cite{Myr15,Hos18}, $f(G)$ model of the mimetic
gravity~\cite{Ast15}, Horndeski mimetic gravity~\cite{Co16,Arr15},
unimodular $f(R)$ mimetic gravity~\cite{Odi16}, $f(R)$ theories in
the mimetic model~\cite{Noj14,Ast16,Noj16} and Galileon mimetic
gravity~\cite{Hag14}. The (in)stability issue in the mimetic gravity
is an important subject that has attracted a lot of attentions. In
fact, the authors seek for the mimetic models which are free of the
ghost and gradient
instabilities~\cite{Ij16,Cap14,Mir14,Mal14,Ram15,Lan15,Ram16,Arr16,Ac16,Hir17,Cai17,Tak17,Yos17,Gan18,Lan18}.
In this regard, the authors of Ref.~\cite{Zhe17} have shown that by
considering the direct coupling between the curvature of the
space-time and the higher derivatives of the mimetic field, it is
possible to have an instabilities-free mimetic model in some ranges
of the parameter space. Also, in Ref.~\cite{Gan20} the authors have
studied the Higher-derivatives Lagrangian model to show the
possibility of evading from instabilities. However, in
Ref.~\cite{Noz19} it has been shown that, considering a DBI mimetic
gravity model is one way to overcome the ghost and gradient
instabilities in the mimetic gravity.

In this paper, we assume the tachyon field to be the mimetic field,
study the inflation and perturbations in this model and compare the
results with observational data. From planck2018 data, we have some
constraints on the perturbation and non-gaussianity  parameters by
which we can explore the observational viability of the inflation
models. In fact, by assuming $\Lambda CDM+r+\frac{dn_{s}}{d\ln k}$
model, Planck2018 TT, TE, EE+lowE+lensing+BAO+BK14 data gives the
constraint on the scalar spectral index as $n_{s}=0.9658\pm0.0038$
and the constraint on the tensor-to-scalar ratio as $r<0.072$,
respectively~\cite{pl18a,pl18b}. The constraint on the tensor
spectral index, implied by Planck2018 TT, TE, EE
+lowE+lensing+BK14+BAO+LIGO and Virgo2016 data is
$-0.62<n_{T}<0.53$~\cite{pl18a,pl18b}. Also, Planck2018 TTT, EEE,
TTE and EET data gives the constraint on the equilateral amplitude
of the non-gaussianity as $f^{equil}=-26\pm 47$~\cite{pl19}. By
numerical studying of these parameters in our inflation model and
compare the results with released data, it is possible to constraint
the model's parameters observationally.

The paper is organized as follows: In section 2, we study the
mimetic tachyon model and obtain the main equations of the model. In
section 3, we consider both the linear and non-linear perturbations
and find perturbation and non-linear parameters in the mimetic
tachyon model. In section 4, we reconstruct the model in terms of
the e-folds number. The power-law inflation in the mimetic tachyon
model is studied in section 5. In this section, we show this model
is free of instabilities. We also find the perturbation and
non-gaussianity parameters in terms of the model's parameter. We
perform a numerical analysis on the model and compare the results
with several observational data sets to obtain some constraints on
the model's parameter space. In section 6, we study the intermediate
inflation in the mimetic tachyon model. In this section also, by
performing a numerical analysis, we explore the (in)stability issue
and the observational viability of the model. In section 7, we
present a summary of the paper.

\section{Mimetic Tachyon Model}

We consider the following action for the mimetic tachyon model in
the lagrange multiplier approach
\begin{eqnarray}
\label{eq3} S=\int
d^{4}x\sqrt{-g}\Bigg[\frac{1}{2\kappa^{2}}R-V(\phi)\,\sqrt{1-2\,\alpha\,X}\nonumber\\
+\lambda(g^{\mu\nu}\partial_{\mu}\phi\,\partial_{\nu}\phi+1) \Bigg],
\end{eqnarray}
where, $R$ is the Ricci scalar, $V(\phi)$ presents potential of the
tachyon field, $\alpha$ is the constant warp factor. Also,
$X=-\frac{1}{2}\,\partial_{\mu}\phi,\partial^{\mu}\phi$ and the
parameter $\lambda$ is a lagrange multiplier, entering the mimetic
constraint (\ref{eq2}) in the action.

Note that, in the Lagrangian formalism generally one is not allowed
to impose the constraints on the action from the beginning. This is
an important point in the lagrangian formalism. We should first
obtain the main equations of motions and then impose the
constraints, as it has been done in paper~\cite{Noz19}. In this
regard, we show that the mimetic tachyon action (\ref{eq3}) leads to
nonzero sound speed providing the propagating curvature
perturbation. The Einstein's field equations in the mimetic tachyon
model are obtained by varying action (\ref{eq3}) with respect to the
metric
\begin{eqnarray}
\label{eq4}
G_{\mu\nu}=\kappa^{2}\Bigg[-g_{\mu\nu}V(\phi)\sqrt{1-2\,\alpha
X}+\frac{\alpha\,
V(\phi)\partial_{\mu}\phi\,\partial_{\nu}\phi}{\sqrt{1-2\,\alpha X}}
\nonumber\\+g_{\mu\nu}\,\lambda\Big(g^{\mu\nu}\,\partial_{\mu}\phi\,\partial_{\nu}\phi+1\Big)-2\lambda\,\partial_{\mu}\phi\,\partial_{\nu}\phi
\Bigg].\hspace{0.5cm}
\end{eqnarray}
In the flat FRW background with the metric
\begin{equation}
\label{eq5} ds^{2}=-dt^{2}+a^{2}(t)\delta_{ij}dx^{i}dx^{j}\,,
\end{equation}
and from the field equations (\ref{eq4}), we obtain the following
Friedmann equations
\begin{eqnarray}
\label{eq6}
3H^{2}=\kappa^{2}\Bigg[\frac{V}{\sqrt{1-\alpha\,\dot{\phi}^{2}}}-\lambda\Big(1+\dot{\phi}^{2}\Big)\Bigg]\,,
\end{eqnarray}

\begin{eqnarray}
\label{eq7}
2\dot{H}+3H^{2}=\kappa^{2}\Bigg[V\,\sqrt{1-\alpha\,\dot{\phi}^{2}}+\lambda\Big(\dot{\phi}^{2}-1\Big)\Bigg]\,.
\end{eqnarray}
Variation of the action (\ref{eq3}) with respect to the tachyon
field gives the following equation of motion in the mimetic tachyon
model
\begin{equation}
\label{eq8}\frac{\alpha\,\ddot{\phi}}{1-\alpha\,\dot{\phi}^{2}}+3\,\alpha\,H\dot{\phi}
-2\lambda\Big(\ddot{\phi}+3H\dot{\phi}\Big)+\frac{V'}{V}-\lambda'\Big(1-\dot{\phi}^{2}\Big)=0\,.
\end{equation}
To study the inflation and observational viability of the mimetic
tachyon model, we should obtain the slow-roll parameters in this
model. These parameters are obtained from the following definitions
\begin{equation}
\label{eq9}\epsilon\equiv-\frac{\dot{H}}{H^{2}}\,,\quad
\eta\equiv\frac{1}{H}\frac{d \ln \epsilon}{dt}\,,\quad
s\equiv\frac{1}{H}\frac{d \ln c_{s}}{dt}\,,
\end{equation}
where $c_{s}$ is the sound speed of the primordial perturbations.
This parameter is defined as $c_{s}^{2}=\frac{P_{,X}}{\rho_{,X}}$
where, $P$ is the pressure, $\rho$ is the energy density and the
subscript ``$,X$" demonstrates derivative of the parameter with
respect to $X$. In this regard, the square of the sound speed in the
mimetic tachyon model is given by
\begin{eqnarray}
\label{eq10}c_{s}^{2}=-{\frac {2\,\kappa^{-2}\dot{H} \left(1-
\alpha\,\dot{\phi}^{2} \right)^{\frac{3}{2}} }{\left[2\, \left(
\alpha\,\dot{\phi}^{2}-1 \right) \lambda\,\sqrt
{-\alpha\,\dot{\phi}^{2}+1} +V \,
\alpha\right]\dot{\phi}^{2}}}\nonumber\\
={\frac { \left( \alpha\,\dot{\phi}^{2}-1 \right)  \left( 2\,\lambda
\sqrt {-\alpha\,\dot{\phi}^{2}+1}-V \, \alpha \right) }{2\, \left(
\alpha\,\dot{\phi}^{2}-1 \right) \lambda\,\sqrt
{-\alpha\,\dot{\phi}^{2}+1} +V \, \alpha}}\,.
\end{eqnarray}
If $0<c_{s}^{2}\leq c^2$, the model is free of gradient
instability~\cite{Eli07,Qui17}.

To seek for the observational viability of the mimetic tachyon
model, in the following we study the perturbations in this setup and
obtain the perturbation and non-gaussian parameters in this setup.

\section{Perturbations in the Mimetic Tachyon Model}
In this section, we study the perturbations in our mimetic tachyon
setup in both linear and non-linear level which help us to explore
the model and its viability, in details.

\subsection{Linear perturbation}
We start with the perturbed ADM line element given by
\begin{eqnarray}
\label{eq11} ds^{2}=
-(1+2{\cal{R}})dt^{2}+2a(t)\Upsilon_{i}\,dt\,dx^{i}\nonumber\\
+a^{2}(t)\left[(1-2{\Psi})\delta_{ij}+2{\Theta}_{ij}\right]dx^{i}dx^{j}\,,
\end{eqnarray}
where $\Upsilon^{i}=\delta^{ij}\partial_{j}\Upsilon+v^{i}$. The
vector $v^{i}$ satisfies the condition $v^{i}_{,i}=0$ and also
${\cal{R}}$ and $\Upsilon$ are 3-scalars~\cite{Muk92}. In this
perturbed metric, we have denoted the spatial curvature perturbation
by $\Psi$ and the spatial symmetric and traceless shear 3-tensor by
${\Theta}_{ij}$. Now, we consider just the scalar part of the the
perturbations at the linear level as
\begin{eqnarray}
\label{eq12}
ds^{2}=-(1+2{\cal{R}})dt^{2}+2a(t)\Upsilon_{,i}\,dt\,dx^{i}\nonumber\\
+a^{2}(t)(1-2{\Psi})\delta_{ij}dx^{i}dx^{j}\,,
\end{eqnarray}
written within the uniform-field gauge ($\delta\phi=0$). We can use
the perturbed metric (\ref{eq12}) and expand the action (\ref{eq3})
up to the second order in the perturbations as
\begin{equation}
\label{eq13} S_{2}=\int
dt\,d^{3}x\,a^{3}{\cal{W}}\left[\dot{\Psi}^{2}-\frac{c_{s}^{2}}{a^{2}}(\partial
{\Psi})^{2}\right],
\end{equation}
which is named the quadratic action and where the parameter
$c_{s}^{2}$ is given by equation (\ref{eq10}). Also, the parameter
${\cal{W}}$ is defined as
\begin{eqnarray}
\label{eq14} {\cal{W}}={\frac { \left[2\, \left(
\alpha\,\dot{\phi}^{2}-1 \right) \lambda\,\sqrt
{-\alpha\,\dot{\phi}^{2}+1} +V \, \alpha\right]\dot{\phi}^{2}
}{2\,H^{2} \left(1- \alpha\,\dot{\phi}^{2} \right)^{\frac{3}{2}} }}
\hspace{1cm}\nonumber\\ ={\frac { -\frac{3}{4}\dot{\phi}^{2} \left(
\left( 2\,\alpha\,\dot{\phi}^{2}-2 \right) \lambda\,\sqrt
{-\alpha\,\dot{\phi}^{2}+1}+V\alpha \right) }{ \left(  \left(
-\dot{\phi}^{2}-1 \right) \lambda\,\sqrt {-\alpha\,\dot{\phi}^
{2}+1}+V \right) {\kappa}^{2} \left(
\frac{1}{2}\,\alpha\,\dot{\phi}^{2}-\frac{1}{2} \right) }} \,.\nonumber\\
\end{eqnarray}
One of the perturbation parameters which is constrained by the
observational data, is the scalar spectral index. To obtain this
parameter in the mimetic tachyon model, we use the following
two-point correlation function
\begin{equation}
\label{eq15} \langle
0|{\Psi}(0,\textbf{k}_{1}){\Psi}(0,\textbf{k}_{2})|0\rangle
=(2\pi)^{3}\delta^{(3)}(\textbf{k}_{1}+\textbf{k}_{2})\frac{2\pi^{2}}{k^{3}}{\cal{A}}_{s}\,,
\end{equation}
with the power spectrum defined as
\begin{equation}
\label{eq16}
{\cal{A}}_{s}=\frac{H^{2}}{8\pi^{2}{\cal{W}}c_{s}^{3}}\,.
\end{equation}
Now, it is possible to find the scale dependence of the perturbation
as
\begin{equation}
\label{eq17} n_{s}-1=\frac{d \ln {\cal{A}}_{s}}{d \ln
k}\Bigg|_{c_{s}k=aH}\,.
\end{equation}
The scalar spectral index, $n_{s}$, in terms of the slow-roll
parameters is obtained as
\begin{equation}
\label{eq18} n_{s}=1-2\epsilon-\eta-s\,.
\end{equation}

By writing the 3-tensor ${\Theta}_{ij}$ of the tensor part of the
perturbed metric (\ref{eq11}), in terms of the two polarization
tensors ($\vartheta_{ij}^{(+,\times)}$) as
${\Theta}_{ij}={\Theta}_{+}\vartheta_{ij}^{+}+{\Theta}_{\times}\vartheta_{ij}^{\times}$,
one can obtain the following expression for the second order action
of the tensor mode
\begin{eqnarray}
\label{eq19} S_{T}=\int dt\, d^{3}x\, \frac{a^{3}}{4\kappa^{2}}
\Bigg[\dot{\Theta}_{+}^{2}-\frac{1}{a^{2}}(\partial
{\Theta}_{+})^{2}+\dot{\Theta}_{\times}^{2}\nonumber\\-\frac{1}{a^{2}}(\partial
{\Theta}_{\times})^{2}\Bigg]\,.
\end{eqnarray}
Following the method used in the scalar part, leads to the amplitude
of the tensor perturbations as
\begin{equation}
\label{eq20} {\cal{A}}_{T}=\frac{2\kappa^{2}H^{2}}{\pi^{2}}.
\end{equation}
By using equations (\ref{eq27})-(\ref{eq29}), we find the tensor
spectral index in this setup as
\begin{equation}
\label{eq21} n_{T}=\frac{d \ln {\cal{A}}_{T}}{d \ln k}=-2\epsilon\,.
\end{equation}

Another important perturbation parameter is the tensor-to-scalar
ratio which is defined as
\begin{equation}
\label{eq22}
r=\frac{{\cal{A}}_{T}}{{\cal{A}}_{s}}=16c_{s}\epsilon\,.
\end{equation}
By performing a numerical analysis on $n_{s}$, $n_{T}$ and $r$ and
comparing the results with the observational data, we can find some
constraints on the model's parameter space. After obtaining the
parameters describing the linear perturbations, in the next
subsection, we study the non-linear perturbations to seek for the
non-gaussian feature of the primordial perturbations and more
constraints on the model's parameters.

\subsection{Non-linear Perturbations}

In studying the primordial perturbations, the linear level of the
perturbations gives us no information about the non-gaussian
feature. Therefore, we should go to the non-linear level of the
perturbations and use the three-point correlation function. In this
regard, by expanding the action (\ref{eq3}) up to the third order in
the small perturbations, and introducing the new parameter
${\cal{Z}}$ satisfying
\begin{eqnarray}
\label{eq23} \Upsilon=\frac{{\Psi}}{H}+\kappa^{2}a^{2}{\cal{Z}}\,,
\end{eqnarray}
and
\begin{equation}
\label{eq24}
\partial^{2}{\cal{Z}}={\cal{W}}\dot{\Psi}\,,
\end{equation}
we find the cubic action, up to the leading order in the slow-roll
parameters of the model, as follows
\begin{eqnarray}
\label{eq25} S_{3}=\int dt\, d^{3}x\,\Bigg\{
\Bigg[\frac{3a^{3}}{\kappa^{2}c_{s}^{2}}\,
\Bigg(1-\frac{1}{c_{s}^{2}}\Bigg) \epsilon
\Bigg]{\Psi}\dot{\Psi}^{2} +\Bigg[\frac{a}{\kappa^{2}}\,\nonumber\\
\Bigg(\frac{1}{c_{s}^{2}}-1\Bigg) \epsilon
\Bigg]{\Psi}\,(\partial{\Psi})^{2}+\Bigg[\frac{a^{3}}{\kappa^{2}}\,
\Bigg(\frac{1}{c_{s}^{2}\,H}\Bigg)\nonumber\\
\Bigg(\frac{1}{c_{s}^{2}}-1\Bigg)\epsilon\Bigg]
\dot{\Psi}^{3}-\Bigg[a^{3}\,\frac{2}{c_{s}^{2}}\epsilon\dot{\Psi}
(\partial_{i}{\Psi})(\partial_{i}{\cal{Z}})\Bigg]\Bigg\}\,.
\end{eqnarray}

In the interaction picture, we have the following expression for the
three-point correlation function for the the spatial curvature
perturbation~\cite{Mal03,Che08}
\begin{eqnarray}
\label{eq26} \langle
{\Psi}(\textbf{k}_{1})\,{\Psi}(\textbf{k}_{2})\,{\Psi}(\textbf{k}_{3})\rangle\hspace{4cm} \nonumber\\
=(2\pi)^{3}\delta^{3}(\textbf{k}_{1}+\textbf{k}_{2}+\textbf{k}_{3}){\cal{B}}_{\Psi}(\textbf{k}_{1},\textbf{k}_{2},\textbf{k}_{3})\,,
\end{eqnarray}
where
\begin{equation}
\label{eq27}
{\cal{B}}_{\Psi}(\textbf{k}_{1},\textbf{k}_{2},\textbf{k}_{3})=\frac{(2\pi)^{4}{\cal{A}}_{s}^{2}}{\prod_{i=1}^{3}
k_{i}^{3}}\,
{\cal{E}}_{\Psi}(\textbf{k}_{1},\textbf{k}_{2},\textbf{k}_{3})\,,
\end{equation}
and the power spectrum ${\cal{A}}_{s}^{2}$ is defined by equation
(\ref{eq16}). The parameter ${\cal{E}}_{\Psi}$ is given by
\begin{eqnarray}
\label{eq28} {\cal{E}}_{\Psi}=\Bigg(1-\frac{1}{c_{s}^{2}}\Bigg)
\Bigg[\frac{3}{4}\Bigg(\frac{2\sum_{i>j}k_{i}^{2}\,k_{j}^{2}}{k_{1}+k_{2}+k_{3}}-\frac{\sum_{i\neq
j}k_{i}^{2}\,k_{j}^{3}}{(k_{1}+k_{2}+k_{3})^{2}}\Bigg)\nonumber\\
-\frac{1}{4}\Bigg(
\frac{2\sum_{i>j}k_{i}^{2}\,k_{j}^{2}}{k_{1}+k_{2}+k_{3}}-\frac{\sum_{i\neq
j}k_{i}^{2}\,k_{j}^{3}}{(k_{1}+k_{2}+k_{3})^{2}}+\frac{1}{2}\sum_{i}k_{i}^{3}\Bigg)\nonumber\\
-\frac{3}{2}\Bigg(\frac{\left(k_{1}\,k_{2}\,k_{3}\right)^{2}}{(k_{1}+k_{2}+k_{3})^{3}}\Bigg)\Bigg]\equiv\Bigg(1-\frac{1}{c_{s}^{2}}\Bigg)
\Upsilon\,,\nonumber\\
\end{eqnarray}
where, the expression inside the square bracket is replaced by
$\Upsilon$. By using the parameter ${\cal{E}}_{\Psi}$, the following
so-called ``non-linearity parameter'', measuring the amplitude of
the non-gaussianity, is defined
\begin{equation}
\label{eq29}
f=\frac{10}{3}\frac{{\cal{E}}_{\Psi}}{\sum_{i=1}^{3}k_{i}^{3}}\,.
\end{equation}
The non-linearity parameter depends on the values of the momenta
$k_{1}$, $k_{2}$ and $k_{3}$. Also, the different values of the
momenta lead to different shapes of the primordial non-gaussianity.
For every shape, there is a special configuration of three momenta,
leading to a maximal signal of the amplitude of the non-gaussianity.
In our model, there is a maximal signal in the equilateral
configuration which has been shown in figure 1. Note that, to plot
this figure, we have introduced the parameters
$x_{2}\equiv\frac{k_{2}}{k_{1}}$ and
$x_{3}\equiv\frac{k_{3}}{k_{1}}$ and we see that there is a peak at
$k_{1}=k_{2}=k_{3}$  (also, in Refs.~\cite{Bab04b,Fel13a,Bau12} it
has been shown that in the k-inflation and higher order derivative
models the signal becomes maximal at the equilateral configuration).
In this regard, in the following, we focus on the equilateral
configuration in which we have $k_{1}=k_{2}=k_{3}$~\cite{Bab04b}. In
this limit, we have

\begin{equation}
\label{eq30}
{\cal{E}}_{\Psi}^{equil}=\frac{17}{72}k^3\left(1-\frac{1}{c_{s}^{2}}\right)\,,
\end{equation}
leading to
\begin{equation}
\label{eq31}
f^{equil}=\frac{85}{324}\left(1-\frac{1}{c_{s}^{2}}\right)\,.
\end{equation}
By using this non-linear parameter obtained in the equilateral
configuration, we can study the non-gaussian feature of the
perturbations in our mimetic tachyon setup numerically.

\begin{figure}
\flushleft\leftskip0em{
\includegraphics[width=.5\textwidth,origin=c,angle=0]{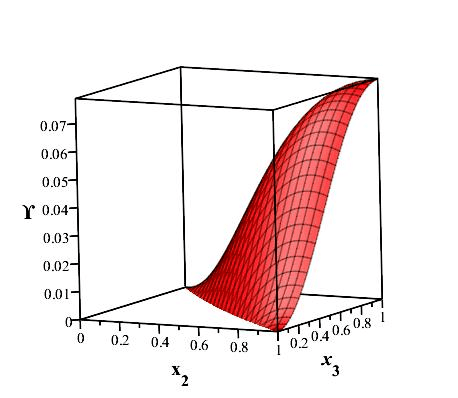}
} \caption{\label{fig7} 3D plot of the function $\Upsilon$ versus
$x_{2}\equiv\frac{k_{2}}{k_{1}}$ and
$x_{3}\equiv\frac{k_{3}}{k_{1}}$. The figure shows that there is a
peak for $\Upsilon$ at $k_{1}=k_{2}=k_{3}$.}
\end{figure}

Note that, to obtain the constraint on $f^{equil}$, we
follow the Planck papers on the
non-gaussianity~\cite{pl19,pl13,pl15}. As has been said in the
mentioned papers, it is possible to use the definition of the
amplitude of the non-gaussianity, sound speed and the scalar
spectral index in every model to find some constraints on the
model's parameters from the observational constraint on $f^{equil}$.
In this way, one can explore the viability of the models. Following
the Planck papers, we use the observational constraints on
$f^{equil}$ to find the constraints on $c_{s}^{2}$ and therefore
model's parameters. Also, we use the constraints on $r-n_{s}$ to
find the constraints on the model's parameters. In this regard, we
can present the prediction of our model for non-gaussianity. In this
regard, we show that it is possible to find some ranges of the
model's parameters which give observationally viable inflation and
primordial perturbation. To this end, we should find the potential
and the lagrange multiplier in our model which we do in the next
sections.

\section{Reconstruction the Model in terms of the e-Folds Number }

One can write the slow-roll parameters, in terms of the e-folds
number defined as
\begin{eqnarray}
\label{eq32}N=\int H\,dt\,.
\end{eqnarray}
To this end, we should first find the potential and lagrange
multiplier in our model. By using equation (\ref{eq7}), and implying
the constraint equation (\ref{eq2}), we find the potential in the
mimetic tachyon model as
\begin{eqnarray}\label{eq33}
V={\frac {2\,H(N) \,H'(N)+3\, H^{2}(N) }{\sqrt
{1-\alpha}{\kappa}^{2}}}\,,
\end{eqnarray}
where we have used a prime to show a derivative of the parameter
with respect to the e-folds number. The lagrange multiplier also is
obtained from equations (\ref{eq6}) and (\ref{eq33}) as follows
\begin{eqnarray}\label{eq34}
\lambda=-{\frac {3\,{H}^{2}(N)\,\alpha+2\,H(N)\,
H'(N)}{2 \left( -1+\alpha \right) {\kappa}^{2}}} \,.\nonumber\\
\end{eqnarray}
After obtaining the potential and lagrange multiplier, we find the
following expression for the sound speed of the mimetic tachyon
model
\begin{eqnarray}\label{eq35}
c_{s}^{2}=-{\frac {2H'(N)\, \left( -1+\alpha \right) ^{2}}{ \left(
4\, \alpha-2 \right) H'(N)+3\,H(N)\, {\alpha}^{2}}} \,.
\end{eqnarray}
Also, the parameter ${\cal{W}}$ is obtained as
\begin{eqnarray}\label{eq36}
{\cal{W}}=\frac {H(N)  \left( 4\,\alpha-2 \right) H'(N) +3\,
H^{2}(N) {\alpha}^{2}}{ 2\left( -1+\alpha \right) ^{2}{\kappa}^{2}{
H}^{2}(N)} \,.
\end{eqnarray}

To find the slow-roll parameters,
following~\cite{Bam14,Odi15}, we introduce a new scalar field
$\varphi$, identified by the number of e-folds $N$. Also, this new
parameter parameterizes the scalar field $\phi$ as
$\phi=\phi(\varphi)$. In this regard, we can write $\dot{\phi}$ as
$\dot{\phi}=\frac{d\phi}{d\varphi}\frac{d\varphi}{dt}=\frac{d\phi}{d\varphi}
H$, which with constraint equation (\ref{eq2}) gives
$\frac{d\phi}{d\varphi}=\frac{1}{H}$. In this way, we have also
$\frac{dV}{d\phi}=\frac{dV}{d\varphi}\frac{d\varphi}{d\phi}=H\frac{dV}{dN}$.
Now, the slow-roll parameters in the mimetic tachyon model and in
terms of the e-folds number are given by
\begin{eqnarray}\label{eq37}
\epsilon=\frac {-H'(N)}{ 2\, H(N) }\,,
\end{eqnarray}
\begin{eqnarray}\label{eq38}
\eta=-\frac {  H(N) H''(N)+\big(H'(N)\big)^{2 } }{ H^(N)\, H'(N)}
\,,
\end{eqnarray}
and
\begin{eqnarray}\label{eq39}
s=\frac {3{\alpha}^{2} \left( H(N) H''(N) - \left( H'(N) \right)
^{2} \right) }{ \left( 8\,\alpha-4 \right) \left( H'(N) \right) ^{2}
+6\,H(N) { \alpha}^{2}H'(N)} \,.
\end{eqnarray}
By using the equations obtained in this section, we can express the
perturbation and non-gaussianity parameters in terms of the Hubble
parameters and therefore e-folds number. In the following, by
adopting some suitable scale factors, we study the mimetic tachyon
model numerically.

\section{Power-Law Inflation in the Mimetic Tachyon Model}

To study the power-law inflation in the mimetic tachyon model, we
use the following scale factor
\begin{eqnarray}
\label{eq40}a=a_{0}\,t^n\,.
\end{eqnarray}
By this scale factor, the Hubble parameter is obtained as follows
\begin{eqnarray}
\label{eq41}H(N)=n\,e^{-\frac{N}{n}}\,.
\end{eqnarray}
Now, from equations (\ref{eq37})-(\ref{eq41}) we find the slow-roll
parameters in the power-law mimetic tachyon model as
\begin{eqnarray}\label{eq42}
\epsilon=\frac {1}{2\,{n}}\,,\quad \eta=\frac {1}{n} \,, \quad s=0
\,.
\end{eqnarray}
Also, the sound speed is obtained as
\begin{eqnarray}\label{eq43}
c_{s}^{2}={\frac { 2\,\left( -1+\alpha \right)
^{2}}{3\,{\alpha}^{2}\,n-4\,\alpha+ 2}} \,,
\end{eqnarray}
and the parameter ${\cal{W}}$ is given by
\begin{eqnarray}\label{eq44}
{\cal{W}}={\frac {3\,{\alpha}^{2}\,n-4\,\alpha+2}{2\,n \left(
-1+\alpha \right) ^{2}{\kappa}^{2}}} \,.
\end{eqnarray}
By using these equations we can find the ranges of the parmeters
$\alpha$ and $n$ leading to gradient and ghost instabilities-free
mimetic tachyon model, corresponding to the constraints
$0<c_{s}^{2}\leq1$ and ${\cal{W}}>0$. The results are shown in
figure 2. As figure shows, the mimetic tachyon model in most ranges
of its parameter space is free of instabilities, making it an
interesting mimetic gravity model.

\begin{figure*}
\flushleft\leftskip0em{
\includegraphics[width=.38\textwidth,origin=c,angle=0]{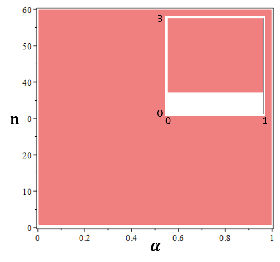}
\hspace{3cm}
\includegraphics[width=.38\textwidth,origin=c,angle=0]{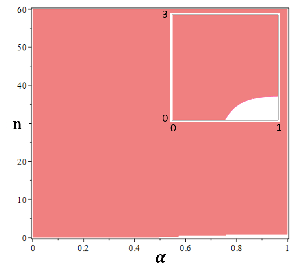}}
\caption{\label{fig2} The coral regions show the ranges of the
model's parameters in the power-law mimetic tachyon model which lead
to $0<c_{s}^{2}\leq1$ (left panel) and ${\cal{W}}>0$ (right panel).}
\end{figure*}

\begin{figure*}
\flushleft\leftskip0em{
\hspace{1cm}\includegraphics[width=.35\textwidth,origin=c,angle=0]{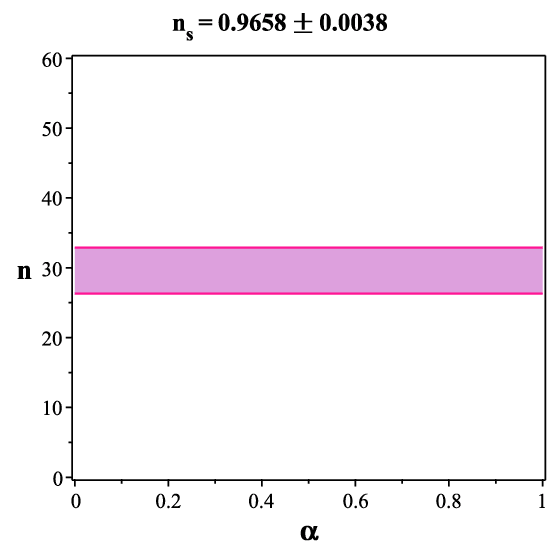}
\hspace{2cm}
\includegraphics[width=.35\textwidth,origin=c,angle=0]{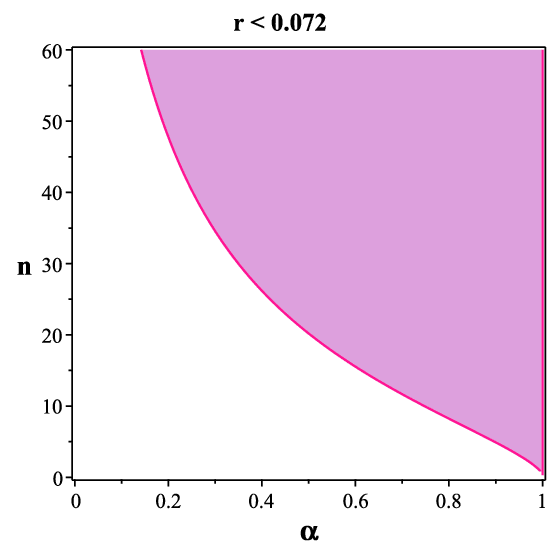}\\
\hspace{5cm} \vspace{1cm}
\includegraphics[width=.35\textwidth,origin=c,angle=0]{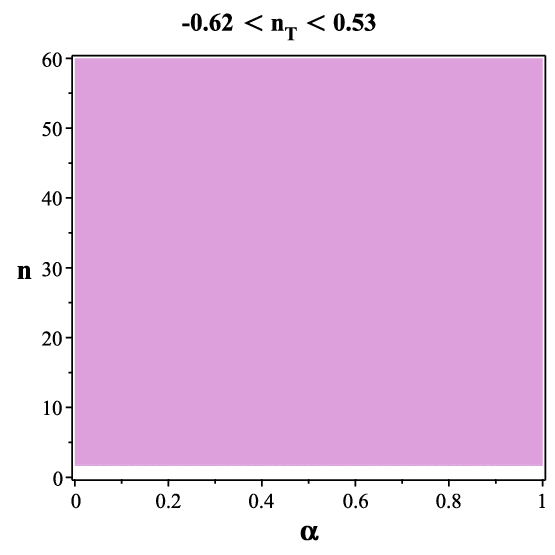}
} \caption{\label{fig3} The plum regions in the upper panels
demonstrate the ranges of the model's parameters in the power-law
mimetic tachyon model leading to the observationally viable values
of the scalar spectral index (left one) and tensor-to-scalar ratio
(right one), which are obtained from Planck2018 TT, TE,
EE+lowE+lensing +BAO +BK14 data. The plum regions in the lower panel
shows the range of the model's parameters leading to the
observationally viable values of the tensor spectral index, which is
obtained from Planck2018 TT, TE, EE +lowE+lensing+BK14+BAO+LIGO and
Virgo2016 data.}
\end{figure*}

We can also study the perturbation parameters numerically to seek
for the observational viability of the mimetic tachyon model. This
gives us more constraints on the model's parameters. By using
equations (\ref{eq18}) and (\ref{eq42}), we can find the scalar
spectral index in terms of the model's parameters and study it
numerically. The left-upper panel of figure 3 shows the ranges of
the parameters $\alpha$ and $n$ which lead to
$n_{s}=0.9658\pm0.0038$, obtained from Planck2018 TT, TE,
EE+lowE+lensing +BAO +BK14 data. The right upper panel of this
figure shows that the ranges of the model's parameters leading to
$r<0.072$, obtained from same data set. The lower panel of figure 3
demonstrates the range of the parameters $\alpha$ and $n$ which
leads to $-0.62<n_{T}<0.53$, obtained from Planck2018 TT, TE, EE
+lowE+lensing+BK14+BAO+LIGO and Virgo2016 data. We have also studied
the behavior of $r-n_{s}$ and $r-n_{T}$ in the background of several
data sets at $68\%$ and $95\%$ CL. The results are shown in figure 4
and 5. The constraints obtained from this numerical analysis are
summarized in table I.

\begin{figure}
\flushleft\leftskip0em{
\includegraphics[width=.5\textwidth,origin=c,angle=0]{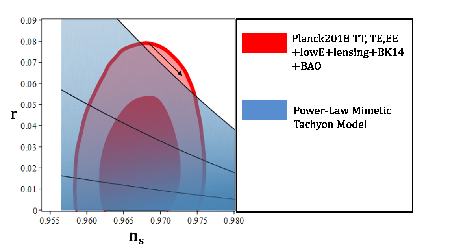}
} \caption{\label{fig4} Tensor-to-scalar ratio versus the scalar
spectral index of the power-law mimetic tachyon model. The black
lines have been drawn to show the behavior of $r-n_{s}$. The
parameter $n$ increases in the direction of the arrow.}
\end{figure}

\begin{figure}
\flushleft\leftskip0em{
\includegraphics[width=.5\textwidth,origin=c,angle=0]{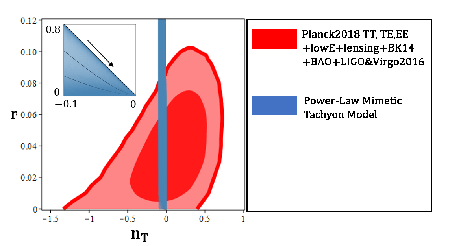}
} \caption{\label{fig5} Tensor-to-scalar ratio versus the tensor
spectral index of the power-law mimetic tachyon model. We have also
zoomed the $r-n_{T}$ plot out to see its evolution clearly. The
parameter $n$ increases in the direction of the arrow.}
\end{figure}

\begin{table*}
\tiny \caption{\small{\label{tab:1} The ranges of the model's
parameters in which the tensor-to-scalar ratio, the scalar spectral
index and the tensor spectral index of the power-law mimetic tachyon
model are consistent with different data sets.}}
\begin{center}
\begin{tabular}{cccccc}
\\ \hline \hline \\ & Planck2018 TT,TE,EE+lowE & Planck2018 TT,TE,EE+lowE&Planck2018 TT,TE,EE+lowE&Planck2018 TT,TE,EE+lowE
\\
& +lensing+BK14+BAO &
+lensing+BK14+BAO&lensing+BK14+BAO&lensing+BK14+BAO
\\
&  & &+LIGO$\&$Virgo2016 &LIGO$\&$Virgo2016
\\
\hline \\$\alpha$& $68\%$ CL & $95\%$ CL &$68\%$ CL & $95\%$ CL
\\
\hline\hline \\  $0.3$& not consistent&$32.4<n<38.5$&$38.1<n< 230.2$
& $30.5<n$\\ \\ \hline
\\$0.5$&$27.9< n<36.4$&$25.3< n <42.1$ &$22.6< n<130.3 $&$18.4<n$
\\ \\ \hline\\
$0.8$&$26.1< n <36.2 $&$24.0< n<41.1$&$9.60< n
<52.4 $ &$7.90<n$\\ \\
\hline \hline
\end{tabular}
\end{center}
\end{table*}

\begin{figure}
\flushleft\leftskip0em{
\includegraphics[width=.37\textwidth,origin=c,angle=0]{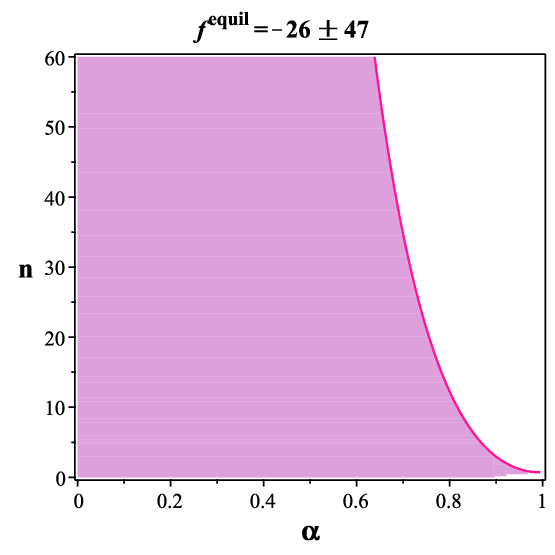}
} \caption{\label{fig6} The plum region shows the ranges of the
model's parameters in the power-law mimetic tachyon model leading to
the observationally viable values of the equilateral amplitude of
the non-gaussianity, which is obtained from Planck2018 TTT, EEE, TTE
and EET data.}
\end{figure}

\begin{table}
\caption{\small{\label{tab:2} The ranges of the model's parameters
in which the equilateral configuration of the non-gaussianity in the
power-law mimetic tachyon model is consistent with the Planck2018
TTT, EEE, TTE and EET data at $68\%$ CL.}}
\begin{center}
\begin{tabular}{cccccc}
\\ \hline \hline   $\alpha=0.3$  & $\alpha=0.5$  &$\alpha=0.8$
\\
\hline\hline \\   $n<1010 $&$
n<186$&$n<12.2$ \\
\hline \hline
\end{tabular}
\end{center}
\end{table}

As mentioned before, the non-gaussian feature of the primordial
perturbations is important issue in studying the inflation models.
Here, we study the equilateral configuration of the non-gaussianity
in comparison with observational data. To numerical study of the
equilateral non-gaussianity, we use equation (\ref{eq31}), where the
sound speed is given by equation (\ref{eq43}). By using the combined
temperature and polarization data analysis at 68$\%$ CL, planck2018
gives the constraint on the equilateral non-gaussianity as
$f^{equil}=-26\pm47$. From this constraint, we have found the ranges
of the model's parameter space leading to the observationally viable
values of the non-gaussianity in the equilateral configuration. The
result is shown in figure 6. This figure shows in some ranges of the
parameter space, we have observationally viable values of the
equilateral non-gaussianity. Also, from Planck2018 TTT, EEE, TTE and
EET data at $68\%$ CL, the constrain on the sound speed is as
$c_{s}^{2}\geq 0.0035$ (this is obtained from the constraint
$f^{equil}=-26\pm47$ released by Planck2018). Table II, shows the
viable ranges of parameter $n$ for some sample values of $\alpha$,
corresponding to this observationally viable range of the square of
the sound speed in the power-law mimetic tachyon model. Figure 7
shows the behavior of the equilateral configuration of the
non-gaussianity versus the sound speed in the background of the
Planck2018 TTT, EEE, TTE and EET data at $68\%$, $95\%$ and $99.7\%$
CL. This figure shows that the equilateral non-gaussianity versus
the sound speed in this model is consistent with observational data.
However, note that, every possible value of the sound speed is not
observationally viable. In fact, according to the equation
(\ref{eq22}), the sound speed is related to the tensor-to-scalar
ratio. The observationally viable values of $r$, set some
constraints on the sound speed of the primordial perturbation.
Figure 8 shows the behavior of the square of the sound speed versus
the tenor-to-scalar ratio in the power-law mimetic tachyon model. To
plot this figure, we have considered the Planck2018 TT, TE,
EE+lowE+lensing +BAO +BK14 data, used in figure 4. From this data
analysis, we have found that for $r>0.44$, the sound speed reach
unity and therefore there would be no non-gaussianity in the model.
To obtain some constraints on the sound speed and the nonlinear
parameter, we use the constraints on $n$, obtained in studying
$r-n_{s}$ behavior in comparison with observational data. The
results are shown in table III. According to our analysis and
considering both Planck2018 TT, TE, EE+lowE+lensing +BAO +BK14 and
Planck2018 TTT, EEE, TTE and EET data sets at $68\%$ CL, the
power-law mimetic tachyon model is observationally viable if
$26.3<n<33.0$ and $0.310<\alpha <0.398$. In these ranges, the model
is instabilities-free and also the perturbation and non-gaussianity
parameters are observationally viable.

\begin{figure}
\flushleft\leftskip0em{
\includegraphics[width=.5\textwidth,origin=c,angle=0]{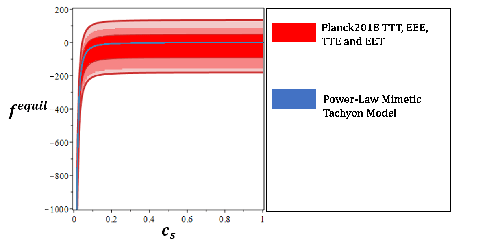}
} \caption{\label{fig7} The amplitude of the non-gaussianity in the
equilateral configuration versus the sound speed in the power-law
mimetic tachyon model.}
\end{figure}

\begin{figure}
\flushleft\leftskip0em{
\includegraphics[width=.37\textwidth,origin=c,angle=0]{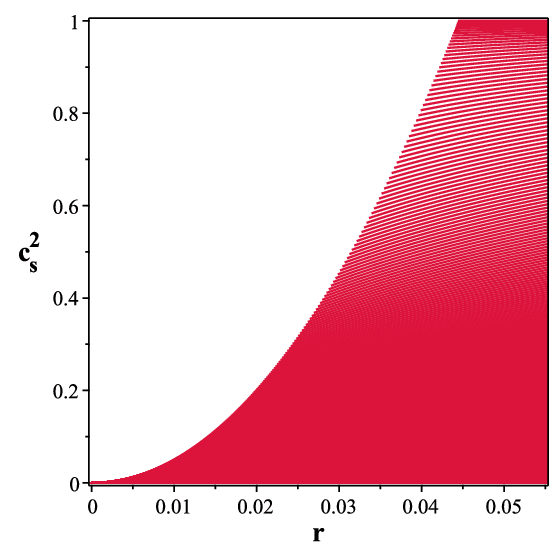}
} \caption{\label{fig8} The square of the sound speed versus the
tensor-to-scalar ratio in the power-law mimetic tachyon model.}
\end{figure}

\begin{table*}
\small \caption{\small{\label{tab:3} The observationally viable
ranges of the several parameters in the power-law mimetic tachyon,
obtained from Planck2018 TT, TE, EE+lowE+lensing+BAO+BK14 data at
$68\%$ CL.}}
\begin{center}
\begin{tabular}{cccccccc}
\\ \hline \hline  \\$\alpha$&& $r$  &&$c_{s}^{2}$  & $f^{equil}$
\\
\hline\hline
\\$0.5$&&$0.029< r <0.044$ &&$0.0173< c_{s}^{2}< 0.0238 $&$-14.0<f^{equil}<-10.7$
\\ \\ \hline\\
$0.8$&&$0.007< r<0.012$&& $0.117\times 10^{-2}< c_{s}^{2}<0.163\times 10^{-2} $ &$-223<f^{equil}<-160$\\ \\
\hline \hline
\end{tabular}
\end{center}
\end{table*}

Note that, the Planck observational data has implied an
upper bound on the tensor-to-scalar ratio. However, there are also
some constraints on the scalar spectral index. When we study
$r-n_{s}$ behavior in our model, we see that as $r$ becomes smaller,
the scalar spectral index becomes larger. In this regard, there
would be both upper and lower limits on the tensor-to-scalar ratio
in our model which beyond those limits the scalar spectral index is
not observationally viable anymore. The existence of these limits on
$r$ leads to the upper and lower limits on $c_{s}^{2}$. Also, we
didn't fix $\epsilon$. This parameter is defined by equation
(\ref{eq42}) which is in terms of the model's parameters. To obtain
the constraints, we have used these equations.

\section{Intermediate Inflation in the Mimetic Tachyon Model}

In this section, we study the intermediate inflation in the mimetic
tachyon model. The intermediate inflation is described by the
following scale factor~\cite{Bar90,Bar93,Bar07}
\begin{eqnarray}
\label{eq45}a=a_{0}\,\exp\left(b\,t^{\beta}\right)\,,
\end{eqnarray}
where, $0<\beta<1$ and $b$ is a constant. The above scale factor of
the intermediate inflation demonstrates that its evolution is faster
than the power law inflation ($a=t^{p}$) but slower than the
standard de Sitter inflation ($a=\exp(Ht)$). By using the scale
factor (\ref{eq45}), we get the following Hubble parameter
\begin{eqnarray}
\label{eq46}H(N)=N \left( {\frac {N}{b}} \right)
^{-{\frac{1}{\beta}}}\beta\,.
\end{eqnarray}
The slow-roll parameters in the intermediate mimetic tachyon model,
obtained from equations (\ref{eq37})-(\ref{eq39}) and (\ref{eq46}),
take the following forms
\begin{eqnarray}\label{eq47}
\epsilon=\frac {\beta-1}{2\,N\,\beta} \,,\,\,\,\quad \eta=\frac
{2-\beta}{N\,\beta}  \,, \hspace{2cm} \nonumber\\  s=\frac
{-3\,\beta\,{\alpha}^{2}}{6\,\beta\,{\alpha}^{2}N+ \left( 8\,
\beta-8 \right) \alpha-4\,\beta+4} \,.\hspace{1cm}
\end{eqnarray}
Also, $c_{s}^{2}$ and ${\cal{W}}$ are obtained as
\begin{eqnarray}\label{eq48}
c_{s}^{2}=\frac { 2\left(1- \beta \right)  \left( -1+\alpha \right)
^{2}}{3 \,\beta\,{\alpha}^{2}N+ \left( 4\,\beta-4 \right)
\alpha-2\,\beta+2} \,,
\end{eqnarray}
\begin{eqnarray}\label{eq49}
{\cal{W}}=\frac {3\,\beta\,{\alpha}^{2}N+ \left( 4\,\beta-4 \right)
\alpha -2\,\beta+2}{2\,\beta\,N \left( -1+\alpha \right)
^{2}{\kappa}^{2}} \,.
\end{eqnarray}
The ranges of $\alpha$ and $n$ leading to gradient and ghost
instabilities-free intermediate mimetic tachyon model are shown in
figure 9. This ranges are corresponding to the constraints
$0<c_{s}^{2}\leq1$ and ${\cal{W}}>0$. From this figure, we find out
that the intermediate mimetic tachyon model too, in some ranges of
its parameter space, is free of gradient and ghost instabilities.

\begin{figure*}
\flushleft\leftskip0em{
\includegraphics[width=.37\textwidth,origin=c,angle=0]{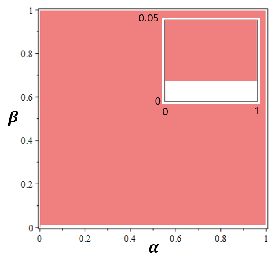}
\hspace{3cm}
\includegraphics[width=.37\textwidth,origin=c,angle=0]{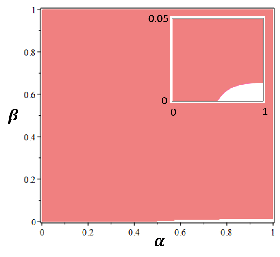}\\
} \caption{\label{fig9} The coral regions show the ranges of the
model's parameters in the intermediate mimetic tachyon model which
lead to $0<c_{s}^{2}\leq1$ (left panel) and ${\cal{W}}>0$ (right
panel).}
\end{figure*}

\begin{figure*}
\flushleft\leftskip0em{
\hspace{1cm}\includegraphics[width=.35\textwidth,origin=c,angle=0]{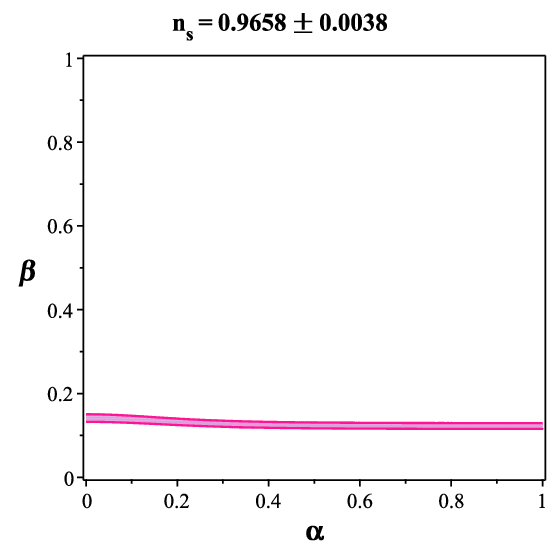}
\hspace{2cm}
\includegraphics[width=.35\textwidth,origin=c,angle=0]{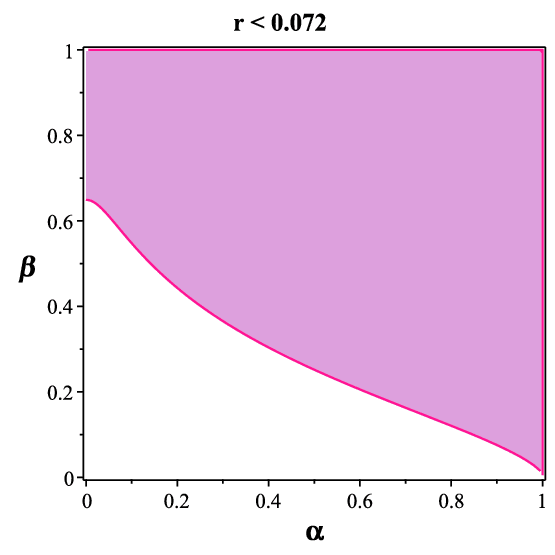}\\
\hspace{5cm} \vspace{0.75cm}
\includegraphics[width=.35\textwidth,origin=c,angle=0]{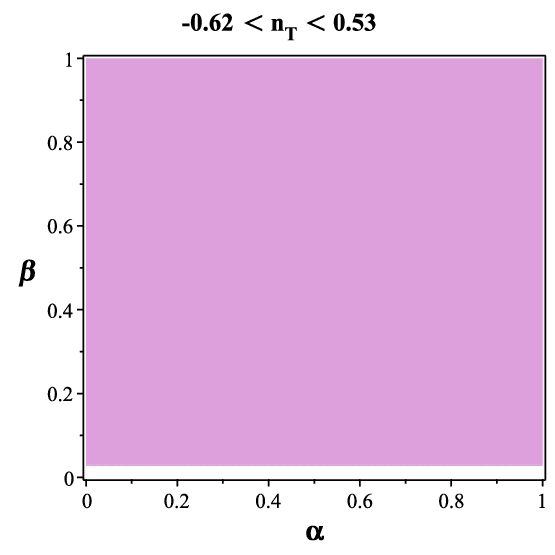}
} \caption{\label{fig10} The plum regions in the upper panels
demonstrate the ranges of the model's parameters in the intermediate
mimetic tachyon model leading to the observationally viable values
of the scalar spectral index (left one) and tensor-to-scalar ratio
(right one), which are obtained from Planck2018 TT, TE,
EE+lowE+lensing +BAO +BK14 data. The plum regions in the lower panel
shows the range of the model's parameters leading to the
observationally viable values of the tensor spectral index, which is
obtained from Planck2018 TT, TE, EE +lowE+lensing+BK14+BAO+LIGO and
Virgo2016 data.}
\end{figure*}

As previous section, we study the perturbation parameters to check
the observational viability of the intermediate mimetic tachyon
model. By substituting equation (\ref{eq47}) in equations
(\ref{eq18}), (\ref{eq21}) and (\ref{eq22}), we obtain the
perturbation parameters in terms of the model's parameter. Now, we
can study the model numerically and compare the results with several
observational data sets. The left-upper panel of figure 10 shows the
ranges of the parameters $\alpha$ and $\beta$ leading to
$n_{s}=0.9658\pm0.0038$. This constraint is obtained from Planck2018
TT, TE, EE+lowE+lensing +BAO +BK14 data. From the same data set, we
have $r<0.072$, leading to the range shown in the right-upper panel
of figure 10. The lower panel of figure 10 demonstrates the range of
the parameters $\alpha$ and $n$ which leads to $-0.62<n_{T}<0.53$,
obtained from Planck2018 TT, TE, EE +lowE+lensing+BK14+BAO+LIGO and
Virgo2016 data. As before, to obtain some constraints on the model's
parameters, we have studied the behavior of $r-n_{s}$ and $r-n_{T}$
in the background of several data sets at $68\%$ and $95\%$ CL. The
results are shown in figure 11 and 12. Table IV shows the
constraints obtained from this numerical analysis. Note that, in the
numerical analysis of this section we adopt $N=60$ and $b=10$.

\begin{figure}
\flushleft\leftskip0em{
\includegraphics[width=.5\textwidth,origin=c,angle=0]{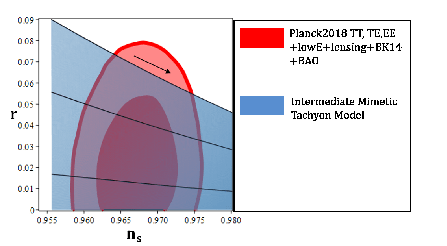}
} \caption{\label{fig11} Tensor-to-scalar ratio versus the scalar
spectral index in the intermediate mimetic tachyon model. The black
lines have been drawn to show the behavior of $r-n_{s}$. The
parameter $n$ increases in the direction of the arrow.}
\end{figure}

\begin{figure}
\flushleft\leftskip0em{
\includegraphics[width=.5\textwidth,origin=c,angle=0]{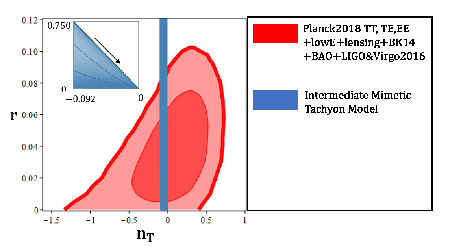}
} \caption{\label{fig12} Tensor-to-scalar ratio versus the tensor
spectral index in the intermediate mimetic tachyon model. We have
also zoomed the $r-n_{T}$ plot out to see its evolution clearly. The
parameter $n$ increases in the direction of the arrow.}
\end{figure}

\begin{table*}
\tiny \caption{\small{\label{tab:4} The ranges of the model's
parameters in which the tensor-to-scalar ratio, the scalar spectral
index and the tensor spectral index in the intermediate mimetic
tachyon model are consistent with different data sets.}}
\begin{center}
\begin{tabular}{cccccc}
\\ \hline \hline \\ & Planck2018 TT,TE,EE+lowE & Planck2018 TT,TE,EE+lowE&Planck2018 TT,TE,EE+lowE&Planck2018 TT,TE,EE+lowE
\\
& +lensing+BK14+BAO &
+lensing+BK14+BAO&lensing+BK14+BAO&lensing+BK14+BAO
\\
&  & &+LIGO$\&$Virgo2016 &LIGO$\&$Virgo2016
\\
\hline \\$\alpha$& $68\%$ CL & $95\%$ CL &$68\%$ CL & $95\%$ CL
\\
\hline\hline \\  $0.8$& not consistent & $0.119<\beta<0.141$&$0.138<\beta<0.455 $ & $0.116<\beta<1$\\ \\
\hline
\\$0.85$&$0.122<\beta<0.133$&$0.114<\beta<0.143$ &$0.115<\beta<0.411 $&$0.097<\beta<1$
\\ \\ \hline\\
$0.9$&$0.117<\beta<0.136$&$0.111<\beta<0.144$&$0.091<\beta<0.332$ &$0.076<\beta<1$\\ \\
\hline \hline
\end{tabular}
\end{center}
\end{table*}

\begin{figure}
\flushleft\leftskip0em{
\includegraphics[width=.37\textwidth,origin=c,angle=0]{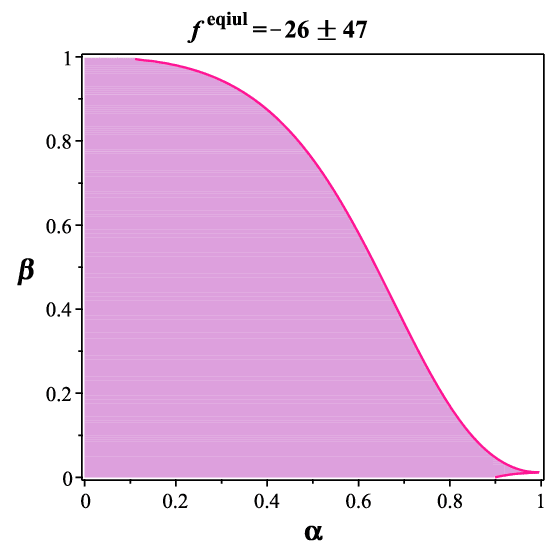}
} \caption{\label{fig13} The plum region shows the ranges of the
model's parameters in the intermediate mimetic tachyon model leading
to the observationally viable values of the equilateral
configuration of the non-gaussianity, which is obtained from
Planck2018 TTT, EEE, TTE and EET data.}
\end{figure}

\begin{table}
\caption{\small{\label{tab:5} The ranges of the model's parameters
in which the equilateral configuration of the non-gaussianity in the
intermediate mimetic tachyon model is consistent with the Planck2018
TTT, EEE, TTE and EET data at $68\%$ CL.}}
\begin{center}
\begin{tabular}{cccccc}
\\ \hline \hline   $\alpha=0.8$  & $\alpha=0.85$  &$\alpha=0.9$
\\
\hline\hline \\   $0.0103<\beta<0.169 $&$
0.0106<\beta<0.096$&$0.0108<\beta<0.046$ \\
\hline \hline
\end{tabular}
\end{center}
\end{table}
Now, we study non-gaussian feature to find more information about
the viability of the intermediate mimetic tachyon model. Here also,
we consider the equilateral configuration of the primordial
non-gaussianity with $k_{1}=k_{2}=k_{3}$. By using equation
(\ref{eq31}), where the sound speed is given by equation
(\ref{eq48}), we can perform a numerical analysis on the equilateral
non-gaussianity. From the constraint $f^{equil}=-26\pm47$, obtained
from the Planck2018 combined temperature and polarization data
analysis at 68$\%$ CL, we have found the observationally viable
range of $\alpha$ and $\beta$, as shown in figure 13. Also, as
mentioned before, from the constraint on $f^{equil}$, we have
$c_{s}^{2}\geq 0.0035$ which gives the viable ranges of parameter
$\beta$ in the intermediate mimetic tachyon model. These ranges, for
some sample values of $\alpha$, are summarized in table V. The
behavior of the equilateral configuration of the non-gaussianity
versus the sound speed in the background of the Planck2018 TTT, EEE,
TTE and EET data at $68\%$, $95\%$ and $99.7\%$ CL is shown in
figure 14. In the intermediate case also, it is necessary to find
the observationally viable values of the sound speed. In this
regard, we use the equation (\ref{eq22}) and the observationally
viable values of $r$, to set some constraints on the sound speed.
The result is shown in figure 15. To plot this figure also, we have
considered the Planck2018 TT, TE, EE+lowE+lensing +BAO +BK14 data,
used in figure 11. From this numerical analysis, we have found that
the constraint on the sound speed in the intermediate mimetic
tachyon model is as $c_{s}^{2}<0.668$ at $68\%$ CL. From the
constraints on $\beta$ in table IV, we have obtained some
constraints on the sound speed and the nonlinear parameter,
summarized in table VI. According to our analysis and based on both
Planck2018 TT, TE, EE+lowE+lensing +BAO +BK14 and Planck2018 TTT,
EEE, TTE and EET data sets at $68\%$, the intermediate mimetic
tachyon model is observationally viable if $0.778<\alpha<0.810$ and
$0.116<\beta<0.130$. In these ranges, the model is free of ghost and
gradient instabilities and also the perturbation and non-gaussianity
parameters are observationally viable.

\begin{figure}
\flushleft\leftskip0em{
\includegraphics[width=.5\textwidth,origin=c,angle=0]{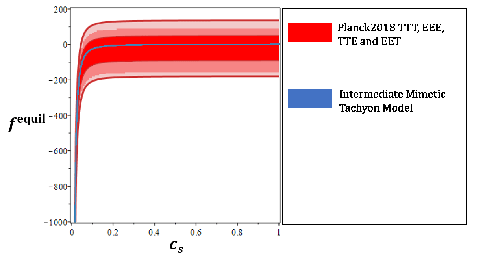}
} \caption{\label{fig14} The amplitude of the non-gaussianity in the
equilateral configuration versus the sound speed in the intermediate
mimetic tachyon model.}
\end{figure}

\begin{figure}
\flushleft\leftskip0em{
\includegraphics[width=.37\textwidth,origin=c,angle=0]{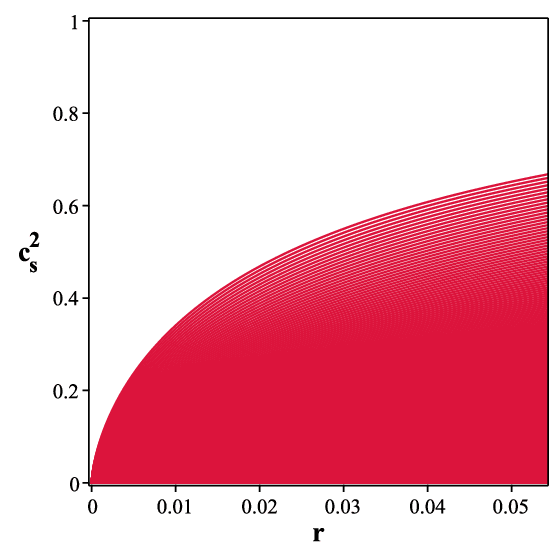}
} \caption{\label{fig15} The square of the sound speed versus the
tensor-to-scalar ratio in the intermediate mimetic tachyon model.}
\end{figure}

\begin{table*}
\small \caption{\small{\label{tab:6} The observationally viable
ranges of the several parameters in the intermediate mimetic
tachyon, obtained from Planck2018 TT, TE, EE+lowE+lensing +BAO +BK14
data at $68\%$ CL.}}
\begin{center}
\begin{tabular}{ccccccc}
\\ \hline \hline  \\$\alpha$&  & $r$&  &$c_{s}^{2}$  && $f^{equil}$
\\
\hline\hline
\\$0.85$&&$0.042< r <0.049$ &&$0.242\times 10^{-2}< c_{s}^{2}< 0.269\times 10^{-2} $&&$-107<f^{equil}<-96.9$
\\ \\ \hline\\
$0.9$&&$0.025< r<0.033$&& $0.936\times 10^{-3}< c_{s}^{2}<0.112\times 10^{-2} $ &&$-279<f^{equil}<-232$\\ \\
\hline \hline
\end{tabular}
\end{center}
\end{table*}

\section{Summary and Conclusion}

In this paper we have studied the tachyon model in the context of
the mimetic gravity and lagrange multiplier approach. We have
assumed that the scalar field in the tachyon model is a mimetic
field. In this regard, we have obtained the Einstein's field
equations, corresponding Friedmann equations and the equation of
motion. After that, we have studied both the linear and non-linear
perturbations in the mimetic tachyon model and found perturbations
and non-gaussianity parameters in terms of the potential of the
mimetic tachyon field and the lagrange multiplier. Then, we have
constructed the model in terms of the Hubble parameter and e-folds
number. This reconstruction has prepared us to study the mimetic
tachyon model for two types of the inflation: power-law and
intermediate inflation.

By adopting the power-law scale factor $a=a_{0}\,t^n$, we have
obtained the slow-roll parameters, sound speed and ${\cal{W}}$ in
terms of $n$ and $\alpha$ (the constant warp factor). By performing
a numerical analysis on these parameters, we have shown that the
power-law mimetic tachyon model in some ranges of its parameter
space is free of gradient and ghost instabilities (corresponding to
$0<c_{s}^{2}\leq 1$ and ${\cal{W}}$, respectively). We have also
studied the perturbation parameters $n_{s}$, $r$ and $n_{T}$
numerically and compared the results with Planck2018 TT, TE,
EE+lowE+lensing +BAO +BK14 and Planck2018 TT, TE, EE
+lowE+lensing+BK14+BAO+LIGO and Virgo2016 data sets. In this regard,
we have obtained some constraints on the parameters $n$ and
$\alpha$, shown in several figures and tables. We have also explored
the non-gaussian feature of the primordial perturbation numerically,
to find more information about the observational viability of the
model. To this end, we have considered the equilateral configuration
of the non-gaussianity with $k_{1}=k_{2}=k_{3}$. By using the
observational constraint on the equilateral amplitude of the
non-gaussianity, we have obtained the ranges of the parameters $n$
and $\alpha$ leading to the viable values of $f^{equil}$. Then, we
have used the relation between the equilateral non-gaussianity and
the sound speed and also the relation between the sound speed and
the tensor-to-scalar ratio, to find some more constraint on the
model's parameter space. Our data analysis, based on both Planck2018
TT, TE, EE+lowE+lensing +BAO +BK14 and Planck2018 TTT, EEE, TTE and
EET data sets at $68\%$ CL, shows that the power-law mimetic tachyon
model is observationally viable and free of instabilities, if
$26.3<n<33.0$ and $0.310<\alpha <0.398$.

We have also checked the mimetic tachyon model with the intermediate
scale factor $a=a_{0}\,\exp\left(b\,t^{\beta}\right)$. By this scale
factor, we have obtained $\epsilon$, $\eta$, $s$, $c_{s}^{2}$ and
${\cal{W}}$ in terms of the intermediate parameters and warp factor.
We have analyzed the parameters $c_{s}^{2}$ and ${\cal{W}}$
numerically and have shown that the intermediate mimetic tachyon
model in some ranges of the model's parameter space is
instabilities-free. By performing a numerical analysis on the
perturbation parameters and comparing the results with Planck2018
TT, TE, EE+lowE+lensing +BAO +BK14 and Planck2018 TT, TE, EE
+lowE+lensing+BK14+BAO+LIGO and Virgo2016 data sets, we have found
some constraints on the model's parameters, which are shown in
several figures and tables. A numerical analysis on the non-gaussian
feature of the primordial perturbation in the intermediate mimetic
tachyon model has shown that it is possible to have the
observationally viable values of the equilateral configuration of
the amplitude of the non-gaussianity in this model. In summary,
using both Planck2018 TT, TE, EE+lowE+lensing +BAO +BK14 and
Planck2018 TTT, EEE, TTE and EET data sets at $68\%$ CL, shows that
the intermediate mimetic tachyon model is observationally viable and
free of ghost instabilities, if $0.778<\alpha<0.810$ and
$0.116<\beta<0.130$.\\

{\bf Acknowledgement}\\
We thank the referee for the very insightful comments that have
improved the quality of the paper considerably.\\

\end{document}